\documentclass[showpacs,preprintnumbers,amsmath,amssymb,
superscriptaddress,floatfix]{revtex4}
\usepackage{graphicx}
\usepackage{amsmath}
\usepackage{bm}
\usepackage{mathrsfs}
\usepackage{color}
\usepackage{slashed}
\usepackage{dcolumn}

\newcommand{\be}{\begin{equation}}
\newcommand{\ee}{\end{equation}}
\newcommand{\ba}{\begin{eqnarray}}
\newcommand{\ea}{\end{eqnarray}}
\newcommand{\nn}{\nonumber}

\newcommand{\la}{\langle}
\newcommand{\ra}{\rangle}

\begin{document}

\title{The pseudoscalar meson and baryon octet interaction
with strangeness zero in the unitary coupled-channel
approximation}

\author{Bao-Xi Sun}
\email{sunbx@bjut.edu.cn}
\affiliation{College of Applied Sciences, Beijing University
of Technology, Beijing 100124, China}

\author{Si-Yu Zhao}
\email{syzhao@emails.bjut.edu.cn}
\affiliation{College of Applied Sciences, Beijing University of
Technology, Beijing 100124, China}

\author{Xiang-Yu Wang}
\email{wxyly_007@163.com}
\affiliation{College of Applied Sciences, Beijing University of
Technology, Beijing 100124, China}


\begin{abstract}
The interaction of the pseudoscalar meson and the baryon octet is
investigated by solving the Bethe-Salpeter equation in the unitary
coupled-channel approximation, In addition to the Weinberg-Tomozawa term,  the contribution of the $s-$
and $u-$ channel potentials in the S-wave approximation are taken
into account.
In the sector of isospin $I=1/2$ and strangeness $S=0$, a pole is
detected in the reasonable region on the complex energy
plane of $\sqrt{s}$ in the center of mass frame by analyzing the
behavior of the scattering amplitude, which is higher than the $\eta
N$ threshold and lies on the third Riemann sheet. Thus it can be
regarded as a resonance state and might correspond to the $N(1535)$
particle in the review of the Particle Data Group(PDG).
The coupling constants of this resonance state to the $\pi N$, $\eta
N$, $K \Lambda$ and $K \Sigma$ channels are calculated, and it is
found that this resonance state couples strongly to the hidden strange channels.
Apparently, the hidden strange channels play an important
role in the generation of the resonance state with strangeness zero.
The interaction of the pseudoscalar meson and the baryon octet is
repulsive in the sector of isospin $I=3/2$ and strangeness $S=0$, therefore, no resonance state can be generated dynamically.
\end{abstract}


\maketitle

\section{Introduction}

The pion-nucleon interaction is an interesting topic and has attracted more attentions of the nuclear society in the past decades.
%
%
There are two very closed excited states of the
nucleon in the $S_{11}$ channel, $N(1535)$ and $N(1650)$, which are
difficult to be described within the framework of the constituent
quark model\cite{Kaiser}. However, in the unitary coupled-channel
approximation of the Bethe-Salpeter equation, most of the excited
states of the nucleon are treated as resonance states of the
pseudoscalar meson and the baryon in the $SU(3)$ flavor space, so
are these two particles. In Ref.~\cite{Kaiser}, it is pointed out
that the hidden strange channels of $K\Lambda$ and $K \Sigma$ might
play an important role in the dynamically generation of the
$N(1535)$ particle.

The $N(1535)$ particle is generated dynamically in the unitary
coupled-channel approximation with the final state interaction of a
three-body $N\pi\pi$ channel considered\cite{Inoue}. However, the
inclusion of the $N\pi\pi$ as a final state in the calculation is
complex, especially there are six arbitrary constants in the real
part of the three-body $N\pi\pi$ loop function, and thus it must be
treated as a free function consistent to the experiment.
This work is studied again by including the $\rho N$ and $\pi
\Delta$ channels in a non-relativistic approximation besides the
pseudoscalar meson -baryon octet channels\cite{Garzon}. Actually the
the elastic scattering process of $\rho N\rightarrow \rho N$ mainly
gives a contribution to the generation of the $N(1650)$ resonance
dynamically, just as done in Ref.~\cite{Ramos}. In the processes of
$\rho N \rightarrow \pi N$ and $\rho N \rightarrow \pi \Delta$, the
Kroll-Ruderman term supplies a constant potential and plays a
dominant role, while the $\pi$-exchange potential is trivial and
proportional to the square of the three-momentum of the final state
in the center of mass frame.
Moreover, the structure of the $N(1535)$ and $N(1650)$ particles are
also studied in the unitary coupled-channel approximation in
\cite{Nieves, Doring}, where a loop function of the intermediate
pseudoscalar meson and the baryon in the on-shell approximation is
taken into account when the Bethe-Salpeter equation is solved.

In Ref.~\cite{Bruns}, the $N(1535)$ and $N(1650)$ resonance states
are studied in the unitary coupled-channel approximation with a
Lagrangian of the pseudoscalar meson and the baryon octet up to the
next-to-leading-order term.
By fitting the $S_{11}$ partial wave amplitude with experimental
data up to the energy $\sqrt{s}=1.56$GeV, the resonance state
corresponding to the $N(1535)$ particle is generated dynamically. In
addition, it is amazing that the $N(1650)$ resonance state can also
be produced at higher energies at the same time.

%
%
The property of the $N(1535)$ particle has also been studied by
solving the relativistic Lippmann-Schwinger equation, where the
corresponding Hamiltonian is divided into two parts, a
non-interacting part and an interacting part, and the couplings and
the bare mass of the nucleon are determined by fitting the
experimental data. This method is named as Hamiltonian effective
field theory by the authors\cite{Liuzw}.
Recently, the different partial wave phase shifts are analyzed by
calculating the K-matrix of the pion-nucleon
interaction\cite{Zheng}.
It is more interesting that the internal wave functions of the
$\Delta(1232)$, $N(1535)$ and $N(1650)$ resonance states are
investigated, and it is announced that the $\pi N$, $\eta N$, $K
\Lambda$ and $K \Sigma$ components are negligible in these resonance
states\cite{Sekihara}. It is apparent that the conclusion made in
Ref.~\cite{Sekihara} is inconsistent with the previous principles
due to chiral unitary models.

In this work, the interaction of the pseudoscalar meson and the
baryon octet is studied in the unitary coupled-channel
approximation, and the contribution of the $s-$ and $u-$ channel
potentials in the S-wave approximation is taken into account besides
the Weinberg-Tomozawa contact term. Furthermore, a revised loop function of
the Bethe-Salpeter equation is used in the
calculation\cite{Dongsun}, where the relativistic correction is
included.

By adjusting the subtraction constants for different intermediate
particles of the loop function in the sector of isospin $I=1/2$ and
strangeness $S=0$, a pole at $1518-i46$MeV on the complex energy
plane is detected, which might be a counterpart of the $N(1535)$
particle in the PDG data\cite{PDG}.

This article is organized as follows. In
Section~\ref{sect:framework}, the potential of the pseudoscalar
meson and baryon octet is constructed, where the Weinberg-Tomozawa contact term, the $s-$ channel and $u-$ channel interactions are all taken into
account in the S-wave approximation.
In Section~\ref{sect:BS}, a basic formula on how to solve the
Bethe-Salpeter equation in the unitary coupled-channel approximation is shown.
The cases of isospin $I=1/2$
and $I=3/2$ are discussed in Section~\ref{sect:I12} and
Section~\ref{sect:I32}, respectively. Finally, the summary is given
in Section~\ref{sect:summary}.

\section{Framework}
\label{sect:framework}

The effective Lagrangian of the pseudoscalar meson and the baryon
octet can be written as
\ba \label{eq:piNLargrangian} {\sl L}=\la \bar{B}(i\gamma_\mu D^\mu
- M) B \ra+\frac{D/F}{2}  \la \bar{B} \gamma_\mu \gamma_5
[u^\mu,B]_{\pm}     \ra. \ea
In the above equation, the symbol $\langle...\rangle$ denotes the
trace of matrices in the $SU(3)$ flavor space, and $D^\mu B =
\partial^\mu B + \frac{1}{2} \left[ [u^\dagger, \partial^\mu u], B
\right]$ with $u^2=U=\exp \left( i\frac{\Phi}{f_0} \right)$ and
$u^\mu = iu^\dagger \partial^\mu u - i u \partial^\mu u^\dagger$,
where $f_0$ is the meson decay constant in the chiral limit.

The matrices of the pseudoscalar meson and the baryon octet are
given as follows

\begin{align}
\Phi={}&\sqrt{2}
\begin{pmatrix}
\frac{1}{\sqrt{2}}\pi^{0}+\frac{1}{\sqrt{6}}\eta & \pi^{+} & K^{+}\\
\pi^{-} &  -\frac{1}{\sqrt{2}}\pi^{0}+\frac{1}{\sqrt{6}}\eta & K^{0}\\
K^{-} & \bar{K}^{0} & -\frac{2}{\sqrt{6}}\eta
\end{pmatrix}
 \label{eq:mesons matrices}
\end{align}
and
\begin{align}
B={}&
\begin{pmatrix}
\frac{1}{\sqrt{2}}\Sigma^{0}+\frac{1}{\sqrt{6}}\Lambda & \Sigma^{+} & p\\
\Sigma^{-} &  -\frac{1}{\sqrt{2}}\Sigma^{0}+\frac{1}{\sqrt{6}}\Lambda & n\\
\Xi^{-} & \Xi^{0} & -\frac{2}{\sqrt{6}}\Lambda
\end{pmatrix}.
 \label{eq:baryons matrices}
\end{align}

\begin{figure}[htb]
\begin{center}
\includegraphics[width=0.65\textwidth]{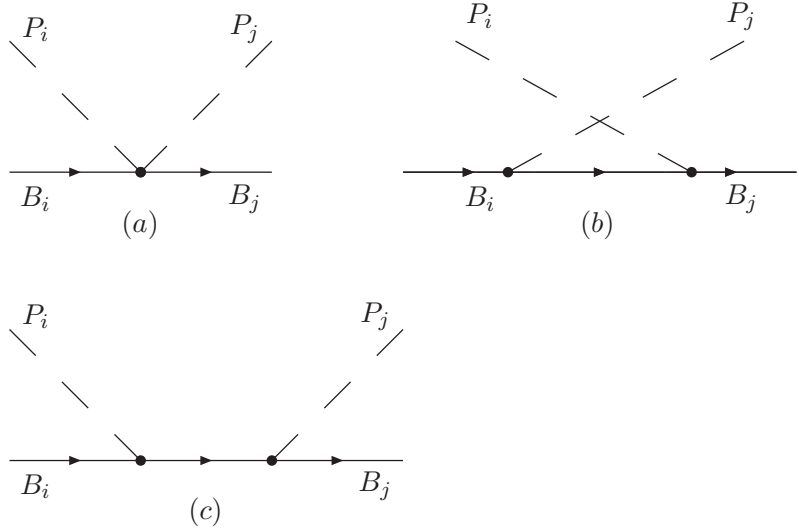}
\end{center}
\caption{Feynman diagrams of the pseudoscalar meson-baryon octet
interaction. $(a)$ contact term, $(b)$ $u-$ channel and $(c)$ $s-$
channel.} \label{fig:Feynman}
\end{figure}

The first term in the Lagrangian in Eq.~(\ref{eq:piNLargrangian})
supplies the contact interaction of the pseudoscalar meson and the
baryon octet, which is usually called as Weinberg-Tomozawa term,
while the other terms which are relevant to the coefficients
$D$ and $F$ give a contribution to the $s-$ and $u-$ channel
interactions, as shown in Fig.~\ref{fig:Feynman}.

According to Feynmann rules, The Weinberg-Tomozawa contact potential
of the pseudoscalar meson and the baryon octet can be written as \be
\label{eq:1903021635} V^{con}_{ij}=-C_{ij}\frac{1}{4f_i
f_j}\bar{U}(p_j,\lambda_j)\gamma_\mu
U(p_i,\lambda_i)(k_i^\mu+k_j^\mu),\ee
where $p_i,p_j(k_i,k_j)$ are the momenta of the initial and final
baryons(mesons), and $\lambda_i, \lambda_j$ denote the spin
orientations of the initial and final baryons, respectively. For low
energies, the three-momenta of the incoming and outgoing mesons can
be neglected, and thus the potential in Eq.~(\ref{eq:1903021635}) is
simplified as
\be
\label{eq:1903021656} V^{con}_{ij}=-C_{ij}\frac{1}{4f_i
f_j}\bar{U}(p_j,\lambda_j)\gamma_0 U(p_i,\lambda_i)(k_i^0+k_j^0).
\ee
Because $U(p_i,\lambda_i)$ and $\bar{U}(p_j,\lambda_j)$ stand for
the wave functions of the initial and final baryons, respectively,
the matrix $\gamma_0$ in Eq.~(\ref{eq:1903021656}) can be replaced
by the unit matrix $I$ at the low energy region, i.e.,
$\gamma_0 \rightarrow I$.
Finally, the Weinberg-Tomozawa contact term of the pseudoscalar
meson and the baryon octet takes the form of
\begin{align}
\label{eq:wtcontact}
V^{con}_{ij}={}&-C_{ij}\frac{1}{4f_i f_j}(2\sqrt{s}-M_{i}-M_{j})
\left(\frac{M_{i}+E}{2M_{i}}\right)^{\frac{1}{2}}
\left(\frac{M_{j}+E^{\prime}}{2M_{j}}\right)^{\frac{1}{2}},
\end{align}
where $\sqrt{s}$ is the total energy of the system, $M_{i}$ and
$M_{j}$ denote the initial and final baryon masses, respectively,
while $E$ and $E^\prime$ stand for the initial and final baryon
energies in the center of mass frame, respectively.
The coefficient $C_{ij}$ for the sector of strangeness zero and
charge zero is listed in Table~\ref{table:coef_S0Q0}, Moreover, we
assume the values of the decay constants are only relevant to the
pseudoscalar meson with $f_{\eta}=1.3f_{\pi}$, $f_K=1.22 f_{\pi}$
and $f_{\pi}=92.4$MeV, as given in Ref.~\cite{Inoue,Bruns}.
\begin{table}[htbp]
\begin{tabular}{c|cccccc}
\hline\hline
 $C_{ij}$          &$K^+ \Sigma^-$ & $K^0 \Sigma^0$ & $K^0 \Lambda$ & $\pi^- p$ & $\pi^0 n$ & $\eta n$    \\
\hline
 $K^+ \Sigma^-$      &$1$ & $-\sqrt{2}$ & $0$ & $0$  & $-\frac{1}{\sqrt{2}}$ & $-\sqrt{\frac{3}{2}}$  \\
 $K^0 \Sigma^0$      &$$ & $0$ & $0$ & $-\frac{1}{\sqrt{2}}$  & $-\frac{1}{2}$ & $\frac{\sqrt{3}}{2}$  \\
$K^0 \Lambda$        &$$ & $$ & $0$ & $-\sqrt{\frac{3}{2}}$  & $\frac{\sqrt{3}}{2}$ & $-\frac{3}{2}$  \\
$\pi^- p$            &$$ & $$ & $$ & $1$  & $-\sqrt{2}$ & $0$  \\
$\pi^0 n$            &$$ & $$ & $$ & $$  & $0$ & $0$    \\
$\eta n$             &$$ & $$ & $$ & $$  & $$ & $0$    \\
 \hline \hline
\end{tabular}
\caption{The coefficients $C_{ij}$ in the pseudoscalar meson and
baryon octet interaction with strangeness $S=0$ and charge $Q=0$,
$C_{ji}=C_{ij}$.} \label{table:coef_S0Q0}
\end{table}

The second term in Eq.~(\ref{eq:piNLargrangian}) supplies
antibaryon-baryon-meson vertices, and can be rewritten as
\be
\label{eq:1812281811}
L= A_{lmn}\bar{N}_l \gamma_\mu \gamma_5 \partial^\mu M_m N_n,
\ee
with $N=\{\Sigma^+, \Sigma^-, \Sigma^0, p, \Xi^-, n, \Xi^0,
\Lambda \}$ and $M=\{\pi^+, \pi^-, \pi^0, K^+, K^-, K^0, \bar{K}^0,
\eta \}$.

The coefficient $A_{lmn}$ in Eq.~(\ref{eq:1812281811}) takes the
form of
\be \label{eq:uschannelcoeff} A_{lmn}=-\frac{1}{2f_0}
 \left[(D+F)C_{lmn}+(D-F)C_{lnm} \right] \ee
where
\be C_{lmn}=\frac{1}{2} \sum_{i,j,k=1}^8 X^\dagger_{il} X_{jm}
X_{kn} \la \lambda_i \lambda_j \lambda_k \ra, \ee
with $\lambda$ the matrix of the SU(3) generator and
\begin{eqnarray}
X =\frac{1}{\sqrt{2}} \left( \begin{array}{cccccccc}
1  & 1 & 0 & 0  & 0 & 0 & 0 & 0 \\
i & -i & 0 & 0  & 0 & 0 & 0 & 0 \\
0 & 0 & \sqrt{2} & 0  & 0 & 0 & 0 & 0 \\
0 & 0 & 0 & 1  & 1 & 0 & 0 & 0 \\
0 & 0 & 0 & i  & -i & 0 & 0 & 0 \\
0 & 0 & 0 & 0  & 0 & 1 & 1 & 0 \\
0 & 0 & 0 & 0  & 0 & i & -i & 0 \\
0 & 0 & 0 & 0  & 0 & 0 & 0 & \sqrt{2} \\
\end{array}\right).
\end{eqnarray}
Thus the $s-$ and $u-$ channel interaction of the pseudoscalar meson
and the baryon octet can be constructed according to the vertices in
Eq.~(\ref{eq:1812281811}).

If the three-momenta of the incoming and outgoing particles are
neglected in the calculation, the $s-$ channel potential of the
pseudoscalar meson and the baryon octet can be written as \ba
\label{eq:schannel} V^s_{ij} &\approx&A A^\prime
\frac{\left(\sqrt{s}-E \right) \left(\sqrt{s}-E^\prime
\right)}{\sqrt{s}+M}, \ea approximately, where $M$ denotes the mass
of the intermediate baryon, $A$ and $A^\prime$ represent the
coefficients depicted in Eq.~(\ref{eq:uschannelcoeff}),
respectively.

Similarly, the $u-$ channel potential can be obtained as \ba
\label{eq:1903011616} V^u_{ij}
&\approx& A A^\prime
\frac{(\sqrt{s}-E)(E+E^\prime-\sqrt{s}-M)(\sqrt{s}-E^\prime)}{u-M^2},
\ea with the Mandelstam variable $u=(p_i-k_j)^2$.

In the calculation of Eqs.~(\ref{eq:schannel}) and
(\ref{eq:1903011616}), a physical baryon mass is adopted so as to
obtain the $s$-channel and $u$-channel interaction potentials of the
pseudoscalar meson and the baryon octet. The mass renormalization of
baryons has be assumed to be accomplished before the tree-level
diagrams in the interaction of the pseudoscalar meson and the baryon
octet are studied. In the chiral unitary model, the loop function of
the intermediate pseudoscalar meson and baryon is considered in an
on-shell approximation when the Bethe-Salpeter equation is solved,
which will be iterated in Sect.~\ref{sect:BS}, so that the whole
interaction chain is taken into account without a cutoff. Therefore,
we can examine whether a resonance state can be generated
dynamically or not.

The Weinberg-Tomozawa term and the $s-$ channel potential of the
pseudoscalar meson and the baryon octet are only related to the
Mandelstam variable $s$, therefore, they only give a contribution to
the S-wave amplitude in the scattering process of the pseudoscalar
meson and the baryon octet.

Since a function can be expanded with the Legendre polynomials,
i.e.,
\be f(x)=\sum_{n=0}^{+\infty} c_n P_n(x), \ee
with $P_n(x)$ the $n$th Legendre polynomial and the coefficient
\be \label{eq:expansion}
 c_n= \frac{2n+1}{2} \int_{-1}^{1} f(x)
P_n(x) dx.
 \ee
In the S-wave approximation, only the coefficient $c_0$ is necessary
to be considered.

the denominator $u-M^2$ in Eq.~(\ref{eq:1903011616}) can be written
as
\ba
u-M^2&=&M_i^2+m_j^2-M^2-2(p_i^0 k_j^0- \vec{p}_i \cdot \vec{k}_j)
\nn \\
&=&(M_i^2+m_j^2-M^2-2p_i^0 k_j^0)\cdot \left(1- \frac{2|\vec{p}_i|
|\vec{k}_j| \cos \theta }{M_i^2+m_j^2-M^2-2p_i^0 k_j^0} \right),
\ea
where $\theta$ is the angle between the three-momenta of incoming
and outgoing mesons, and $\vec{p}_i(\vec{k}_j)$ and $M_i(m_j)$ are
the three-momentum in the center of mass frame and the mass of the
initial baryon (final meson), respectively. Supposing
$\alpha=\frac{2 |\vec{p}_i| \vec{k}_j| }{M_i^2+m_j^2-M^2-2E
(\sqrt{s}-E^\prime)}$ and $x=\cos \theta$, we can obtain
\be \frac{1}{2}~\int^1_{-1} \frac{1}{1-\alpha x} dx =
-\frac{1}{2}~\frac{1}{\alpha}\ln \left(\frac{1-\alpha}{1+\alpha}
\right).
\ee
Thus the $u-$ channel potential of the pseudoscalar meson and the
baryon octet in the S-wave approximation can be calculated easily
\ba
\label{eq:Vuswave}
V^u_{ij}(S)
&=&A A^\prime
\frac{(\sqrt{s}-E)(E+E^\prime-\sqrt{s}-M)(\sqrt{s}-E^\prime)}
{M_i^2+m_j^2-M^2-2E (\sqrt{s}-E^\prime)}\cdot \frac{-1}{2\alpha} \ln
\left(\frac{1-\alpha}{1+\alpha} \right), \ea
Therefore, the S-wave potential of the pseudoscalar meson and the
baryon octet can be written as
\be \label{eq:1903032134} V_{ij}=V_{ij}^{con}+V_{ij}^s+V_{ij}^u(S).
\ee

\section{Bethe-Salpeter equation}
\label{sect:BS}

The Bethe-Salpeter equation can be expanded as
\ba
\label{eq:1903021709}
T &=&V+VGT \nn \\
 &=&V+VGV+VGVGV+....
\ea
When the Bethe-Salpeter equation in Eq.~(\ref{eq:1903021709}) is
solved, only the on-shell part of the potential $V_{ij}$ in
Eq.~(\ref{eq:1903021656}) gives a contribution to the amplitude of
the pseudoscalar meson and the baryon octet, and the off-shell part
of the potential can be reabsorbed by a suitable renormalization of
the decay constants of mesons $f_i$ and $f_j$.
More detailed discussion can be found in
Refs.~\cite{Oller97,Ramos97}.
Therefore,
if the potential in Eq.~(\ref{eq:1903021656}) is adopted, the second
term $VGV$ in Eq.~(\ref{eq:1903021709}) can be written as
\be
V_{jl}G_{l} V_{li} \sim  \bar{U}(p_j,\lambda_j) G_{l}
 U(p_i,\lambda_i)(k_i^0+k_j^0)^2.
 \ee
If the relativistic kinetic correction of the loop function of the
pseudoscalar meson and the baryon octet is taken into account, the
loop function $G_{l}$ can be written as
\begin{align}
G_{l}={}&i\int\frac{d^{d}q}{(2\pi)^{4}}\frac{\rlap{/}q+M_{l}}
{q^{2}-M_{l}^{2}+i\epsilon}\frac{1}{(P-q)^{2}-m_{l}^{2}+i\epsilon},
 \label{eq:G}
\end{align}
with $P$ the total momentum of the system, $m_{l}$ the meson mass,
and $M_{l}$ the baryon mass, respectively.

The loop function in Eq.~(\ref{eq:G}) can be calculated in the
dimensional regularization (See Appendix 1 of Ref.~\cite{Dongsun}
for details), and thus the loop function takes the form of
\begin{equation}
\begin{aligned}
G_{l}={}&\frac{\gamma_{\mu}
P^{\mu}}{32P^{2}\pi^{2}}\left[(a_{l}+1)(m_{l}^{2}-M_{l}^{2})+(m_{l}^{2}\ln\frac{m_{l}^{2}}{\mu^{2}}-M_{l}^{2}\ln\frac{M_{l}^{2}}{\mu^{2}})\right]\\&
+\left(\frac{\gamma_{\mu}
P^{\mu}[P^{2}+M_{l}^{2}-m_{l}^{2}]}{4P^{2}M_{l}}+\frac{1}{2}\right)G_{l}^{\prime},
\end{aligned} \label{eq:Our G}
\end{equation}
where $a_l$ is the subtraction constant and $\mu$ is the
regularization scale, and $G_{l}^{\prime}$ is the loop function in
Ref.~\cite{Oller},
\begin{eqnarray}
G^\prime_{l}(s) &=& \frac{2 M_l}{16 \pi^2} \left\{ a_l(\mu) + \ln
\frac{m_l^2}{\mu^2} + \frac{M_l^2-m_l^2 + s}{2s} \ln
\frac{M_l^2}{m_l^2} + \right. \nonumber \\ & &  \phantom{\frac{2
M}{16 \pi^2}} + \frac{\bar{q}_l}{\sqrt{s}} \left[
\ln(s-(M_l^2-m_l^2)+2\bar{q}_l\sqrt{s})+
\ln(s+(M_l^2-m_l^2)+2\bar{q}_l\sqrt{s}) \right. \nonumber  \\
& & \left. \phantom{\frac{2 M}{16 \pi^2} +
\frac{\bar{q}_l}{\sqrt{s}}} \left. \hspace*{-0.3cm}-
\ln(-s+(M_l^2-m_l^2)+2\bar{q}_l\sqrt{s})-
\ln(-s-(M_l^2-m_l^2)+2\bar{q}_l\sqrt{s}) \right] \right\},
\label{eq:gpropdr}
\end{eqnarray}
with $\bar{q}_l$ the three-momentum of the meson or the baryon in
the center of mass frame.

Since the total three-momentum $\vec{P}=0$ in the center of mass
frame, only the $\gamma_{0} P^{0}$ parts remain in Eq.~(\ref{eq:Our
G}).
Similarly, This matrix $\gamma_{0}$ can be replaced by the unit
matrix $I$ since the $U(p_i,\lambda_i)$ and $\bar{U}(p_j,\lambda_j)$
denote the wave functions of the initial and final baryons,
respectively. Thus the loop function of the intermediate
pseudoscalar meson and baryon octet becomes
\begin{equation}
\begin{aligned}
G_{l}={}&\frac{\sqrt{s}}{32\pi^{2}s}\left[(a_l+1)(m_{l}^{2}-M_{l}^{2})+(m_{l}^{2}ln\frac{m_{l}^{2}}{\mu^{2}}-M_{l}^{2}ln\frac{M_{l}^{2}}{\mu^{2}})\right]\\&
+\left(\frac{s+M_{l}^{2}-m_{l}^{2}}{4M_{l}\sqrt{s}}+\frac{1}{2}\right)G_{l}^{\prime}.
\end{aligned}\label{eq:Our G result}
\end{equation}

When the $s-$ channel and $u-$ channel interaction are supplemented,
the loop function in Eq.~(\ref{eq:Our G result}) is still suitable.
However, the off-shell part of the potential is reabsorbed by a
renormalization, so the decay constants of mesons, the masses of
intermediate baryons all take physical values when the
Bethe-Salpeter equation is solved.

In the calculation of the present work, we make a transition of
\ba \tilde{V}&=&V~\sqrt{M_i M_j}, \nn \\
    \tilde{G}_l&=&G_l/M_l,
\ea
so the scattering amplitude
\be \tilde{T}=[1-\tilde{V}\tilde{G}]^{-1}\tilde{V} \ee becomes
dimensionless.

\section{$I=\frac{1}{2}$ and $S=0$}
\label{sect:I12}

\begin{figure}[!htb]
\centerline{
\includegraphics[width = 0.35\linewidth]{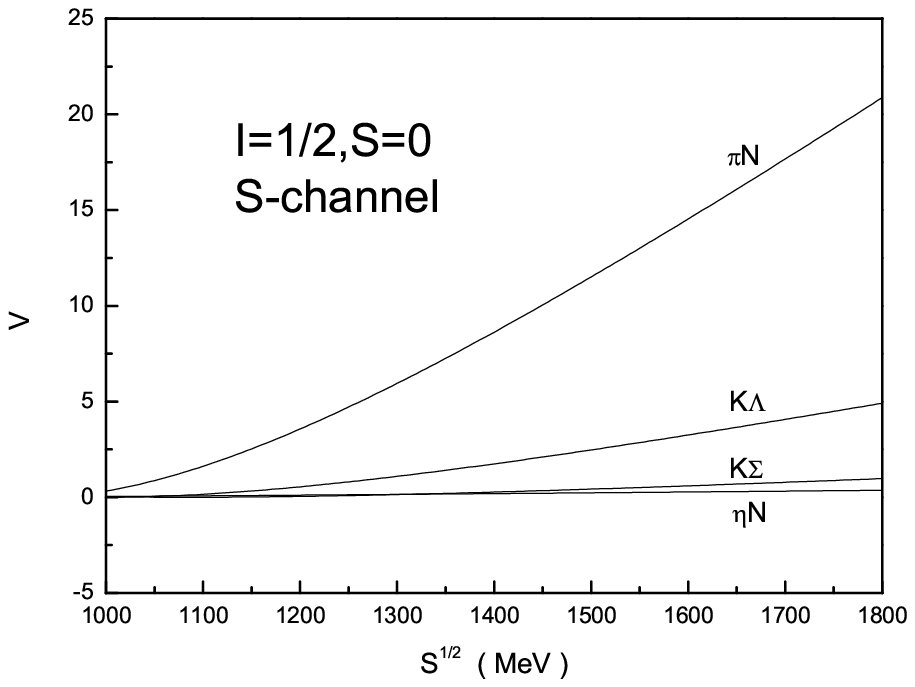}
\includegraphics[width = 0.35\linewidth]{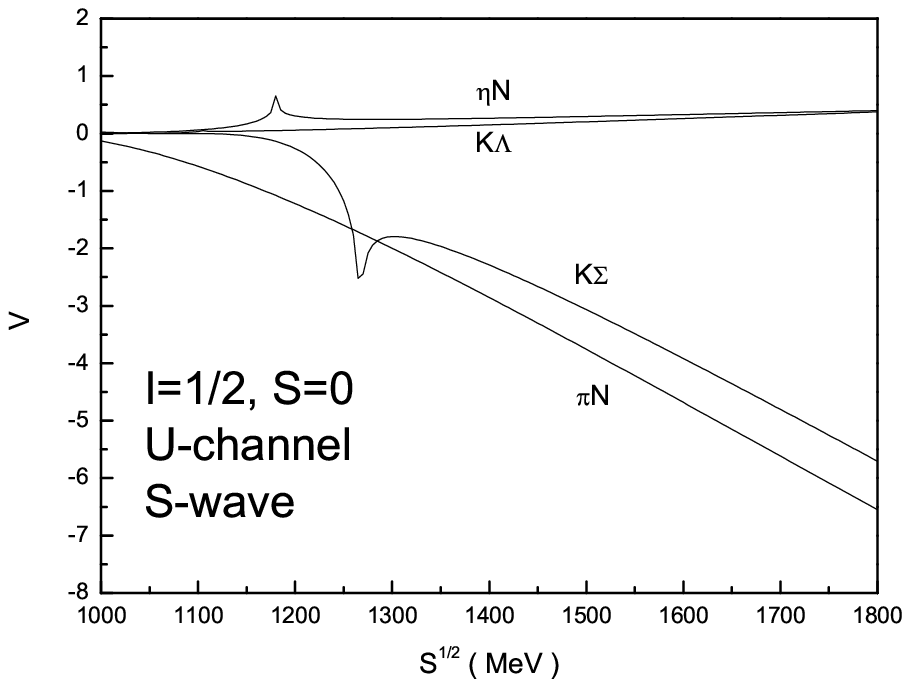}
\includegraphics[width = 0.35\linewidth]{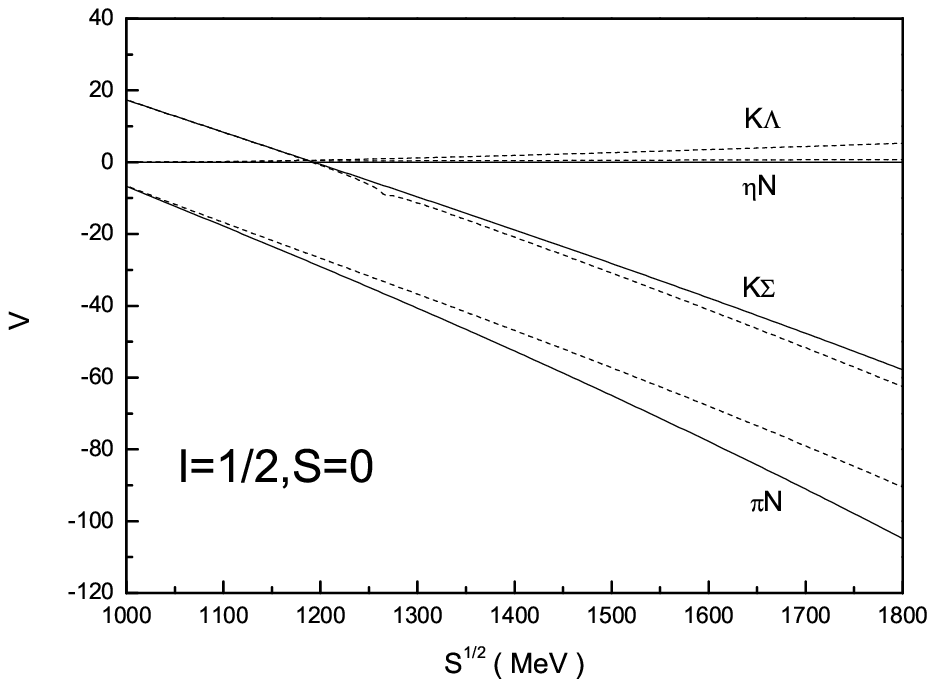}
}
\caption{Potentials of the pseudoscalar meson and the baryon octet
as functions of the total energy of the system $\sqrt{s}$ in the
sector of isospin $I=1/2$ and strangeness $S=0$. Left: $s-$ channel,
Middle: $u-$ channel in the S-wave approximation, Right: The solid
lines denote the Weinberg-Tomozawa contact interaction, while the
dash lines stand for the total S-wave potential in
Eq.~(\ref{eq:1903032134}). }
 \label{fig:schi12s0}
\end{figure}

In the sector of isospin $I=\frac{1}{2}$ and strangeness $S=0$, the
wave function in the isospin space can be written as
\be
| \pi N; \frac{1}{2}, -\frac{1}{2} \ra = -\sqrt{\frac{2}{3}} |\pi^-
p \ra + \sqrt{\frac{1}{3}} | \pi^0 n \ra,
 \ee

\be | \eta N; \frac{1}{2}, -\frac{1}{2}\ra = |\eta n \ra,
 \ee

\be |K \Lambda; \frac{1}{2}, -\frac{1}{2} \ra =|K^0 \Lambda \ra, \ee
and
 \be |K \Sigma; \frac{1}{2}, -\frac{1}{2} \ra =
-\sqrt{\frac{1}{3}} |K^0 \Sigma^0 \ra + \sqrt{\frac{2}{3}} |K^+
\Sigma^- \ra, \ee respectively. Thus the coefficients $C_{ij}$ in
the Weinberg-Tomozawa contact potential of the pseudoscalar meson
and the baryon octet can be obtained in the isospin space, which are
summarized in Table~\ref{table:coef_S0I12}.

\begin{table}[htbp]
\begin{tabular}{c|cccc}
\hline\hline
 $C_{ij}$         & $\pi N$ &  $\eta N$   &$K \Lambda$  &  $K \Sigma$    \\
\hline
$\pi N$            &$2$ & $0$ & $\frac{3}{2}$ & $-\frac{1}{2}$ \\
$\eta N$           &$$ & $0$ & $-\frac{3}{2}$ & $-\frac{3}{2}$ \\
$K \Lambda$        &$$ & $$ &$0$  & $0$   \\
$K \Sigma$         &$$ & $$ & $$ & $2$   \\
\hline \hline
\end{tabular}
\caption{The coefficients $C_{ij}$ in the pseudoscalar meson and
baryon octet interaction with isospin $I=\frac{1}{2}$ and
strangeness $S=0$, $C_{ji}=C_{ij}$.} \label{table:coef_S0I12}
\end{table}

The $s-$ channel, $u-$ channel and Weinberg-Tomozawa contact
potentials of the pseudoscalar meson and baryon octet in the S-wave
approximation are depicted in Figs.~\ref{fig:schi12s0},
respectively. In Fig.~\ref{fig:schi12s0}, it is found that the $\pi
N$ $s-$ channel potential is repulsive and the other $s-$ channel
potential are weaker than the $\pi N$ case,
while the $u-$ channel potentials in the S-wave approximation are
attractive.
Although the curves for $\eta N$ and $K\Sigma$ cases are not smooth
when $\sqrt{s}<1300$MeV, it is far away from the energy region which
we are interested in, and we assume that it would not give an effect
on the pole position of the amplitude in the calculation.
However, the contact interaction originated from the
Weinberg-Tomozawa term is dominant in the pseudoscalar meson and the
baryon octet potential, and the correction from the $s-$ channel
potential and the S-wave $u-$ channel potential is not important.

The total potentials for different pseudoscalar meson and baryon
systems with isospin $I=1/2$ and strangeness $S=0$ are depicted in
the right figure  of Fig.~\ref{fig:schi12s0}.
It shows that the $\pi N$ and $K \Sigma$ potentials are attractive,
while the $\eta N$ and $K \Lambda$ interactions are weak.
Although the $s-$ channel and $u-$ channel potentials are weaker
than the Weinberg-Tomozawa contact interaction in the sector of
isospin $I=1/2$ and strangeness $S=0$, the subtraction constants
must be readjusted in the calculation when the contribution of the
$s-$ and $u-$ channel potentials are taken into account.

According to the PDG data, the $N(1535)$ particle is assume to lie in the region of 
$Re(pole position)=1490\sim1530$MeV, and $-2Im(pole position)=90\sim250$MeV on the complex energy plane of $\sqrt{s}$\cite{PDG}.
When the Bethe-Salpeter equation is solved in the unitary
coupled-channel approximation, we set the regularization scale $\mu
= 630$MeV, just as done in most of works with this method\cite{Garzon,Ramos,Dongsun,Ramos97}. Moreover, all of subtraction constants change from $-3.2$ to $-0.5$ with a step of $0.3$,
and we hope a resonance state can be generated dynamically in the reasonable energy region.
In the previous works, the subtraction constant is usually chosen to be $-2$, which is thought to be a natural value. we change the subtraction constant values in the neighboring region of $-2$ in order to find the influence of different subtraction constant values on the mass and decay width of the resonance state.

\begin{figure}[!htb]
\centerline{
\includegraphics[width = 0.8\linewidth]{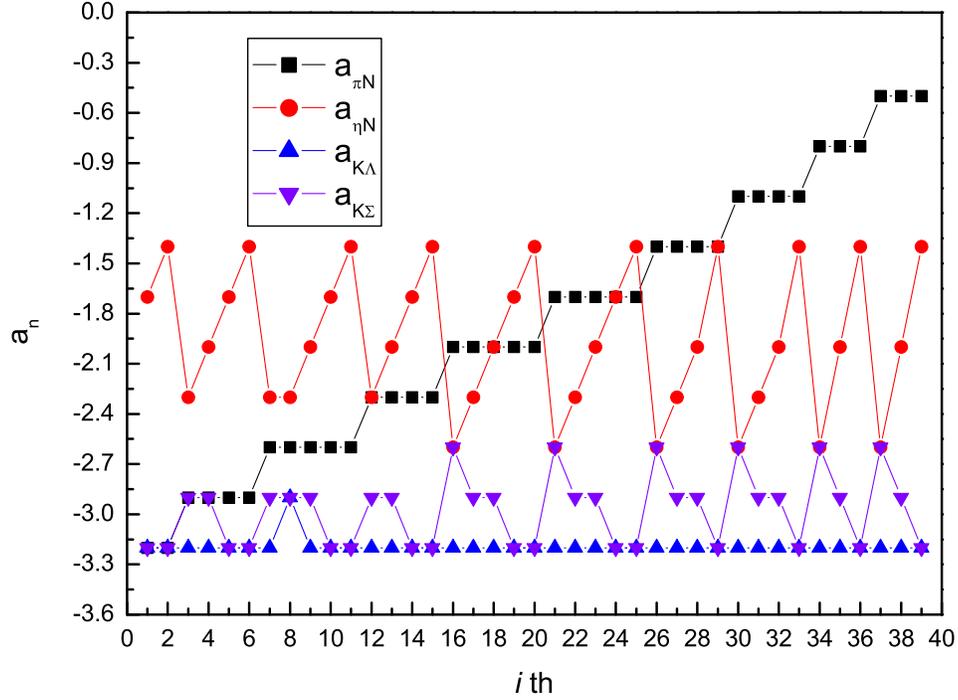}
}
\caption{The values of the subtraction constant $a_{\pi N}$,
$a_{\eta N}$, $a_{K \Lambda}$, $a_{K \Sigma}$ with the regularization
scale $\mu=630$MeV fixed in the loop function in Eq.~(\ref{eq:Our G result}).
 }
 \label{fig:para2019123}
\end{figure}

Altogether we find 39 sets of subtraction constant values is suitable to produce a pole in the energy region constrained by the PDG data, 
which are listed in the Appendix part of this manuscript. At the same time, the pole position and the couplings to $\pi N$, $\eta N$, $K \Lambda$ and $K \Sigma$ are also listed. A resonance state with a mass about $1520$MeV and a decay width about $90$MeV is generated by using 12 sets of subtraction constants, while both the mass and the decay width of the resonance state increase slightly when the other 27 sets of parameters are used in the calculation, respectively.
These 39 sets of subtraction constants are depicted in Fig.~\ref{fig:para2019123}, and it is found that the subtraction constant $a_{\pi N}$ changes from $-3.2$ to $-0.5$ successively. Since the $\pi N$ threshold is far lower than the energy region where the $N(1535)$ particle might be generated dynamically, it is understandable that the pole position is not sensitive to the value of the subtraction constant $a_{\pi N}$.
The changes of the other three subtraction constants $a_{\eta N}$, $a_{K \Lambda}$ and $a_{K \Sigma}$ are not so large. Especially, the subtraction constant $a_{K \Lambda}=-3.2$ in 38 sets of parameters except the eighth set, where it takes the value of $-2.9$. The $K \Lambda$ threshold is close to the energy region where we are interested, thus the subtraction constant $a_{K \Lambda}$ is stable and plays an important role in the generation of the $N(1535)$ particle.

%

\begin{figure}[!htb]
\centerline{
\includegraphics[width = 0.65\linewidth]{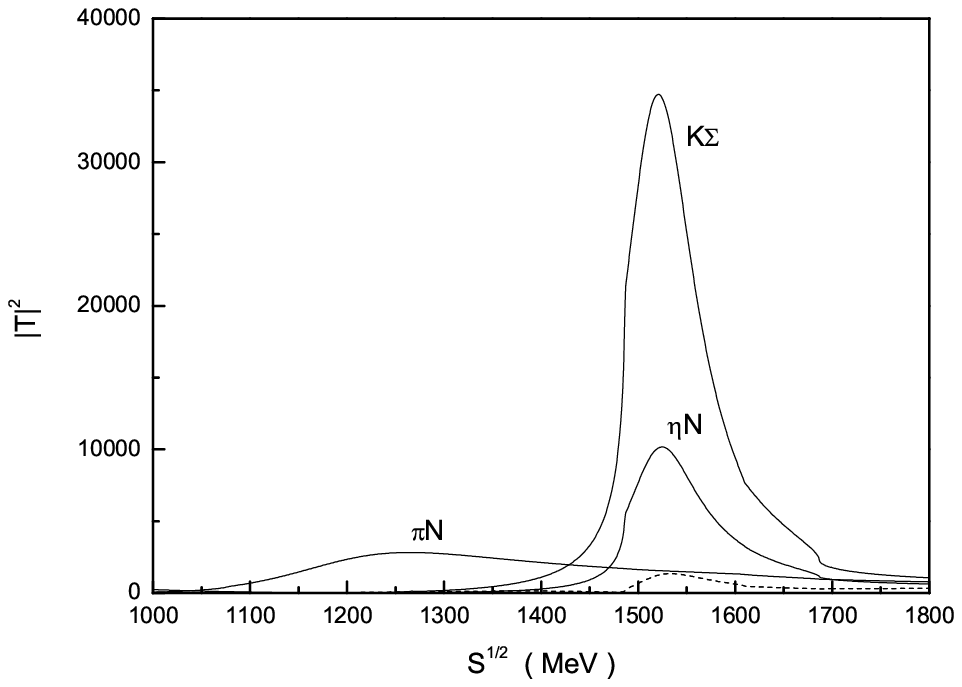}
}
\caption{The squared amplitude $|T|^2$ as a function of the total
energy $\sqrt{s}$ for different channels with isospin $I=1/2$ and
strangeness $S=0$. The cases of $\pi N$, $\eta N$ and $K\Sigma$
channels are labeled on the figure, while the case of the $K
\Lambda$ channel is drawn with the dash line.}
 \label{fig:i12s0swave}
\end{figure}

A pole is generated dynamically at $1518-i46$MeV on the complex
energy plane of $\sqrt{s}$ by solving the Bethe-Salpeter equation in
the unitary coupled-channel approximation with the 19th set of parameters, i.e., $a_{\pi N}=-2.0$, $a_{\eta N}=-1.7$, $a_{K \Lambda}=-3.2$ and $a_{K \Sigma}=-3.2$.
The squared amplitude $|T|^2$ as a function of the total energy
$\sqrt{s}$ for different channels with isospin $I=1/2$ and
strangeness $S=0$ are depicted in Fig.~\ref{fig:i12s0swave}.
The real part of the pole position is higher than the $\eta N$
threshold, and lower than the $K \Lambda$ threshold,
so we assume it might be a resonance state and correspond to the
N(1535) particle in the PDG data.

The couplings of the resonance state to the
$\pi N$, $\eta N$, $K \Lambda$ and $K \Sigma$ channels are also calculated.
and it is found that
it couples strongly to the $\eta N$, $K \Lambda$ and $K \Sigma$
channels, which implies that these channels is important 
in the generation of the $N(1535)$ resonance state. 
If different sets of subtraction constants are chosen in the calculation, the changes of couplings are not significant, as shown in the table of the Appendix part.

In Ref.~\cite{Bruns}, Eq.~(15) indicates that the $N(1535)$ particle
couples more strongly to the $K^{+} \Lambda$ channel, which is
different from the results listed in Table~\ref{table:subtract}.
The different values of the coupling constants might be relevant to
the next-to-leading-order chiral Lagrangian used in
Ref.~\cite{Bruns}, while it is not included in this manuscript.

\section{$I=\frac{3}{2}$ and $S=0$}
\label{sect:I32}

\begin{figure}[!htb]
\centerline{
\includegraphics[width = 0.65\linewidth]{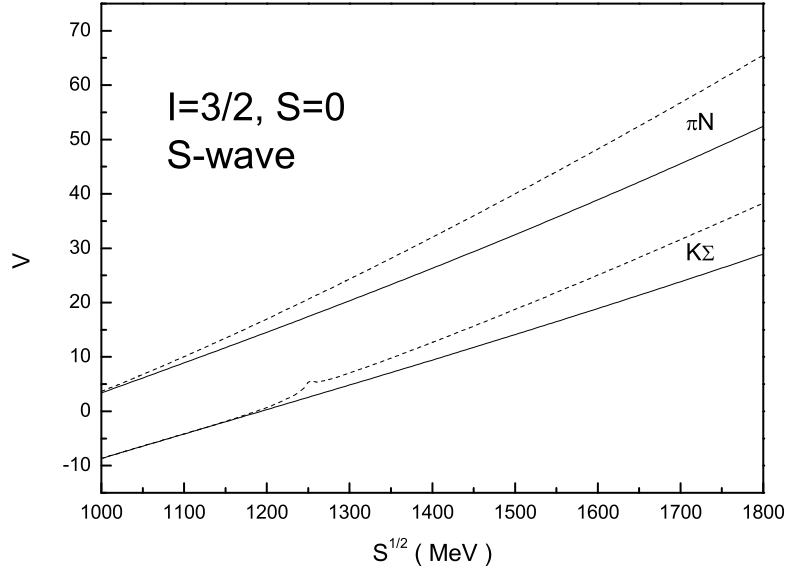}
}
\caption{The potential of the pseudoscalar meson and the baryon
octet as a function of the total energy of the system $\sqrt{s}$ in
the sector of isospin $I=3/2$ and strangeness $S=0$. The solid lines
denote the contact interaction, while the dash lines stand for the
total S-wave potential in Eq.~(\ref{eq:1903032134}).}
\label{fig:tchi32s0}
\end{figure}

The wave functions with isospin $I=3/2$ and strangeness $S=0$ can be
written as
\be \label{eq:1901101}
| \pi N; \frac{3}{2}, -\frac{1}{2} \ra = \sqrt{\frac{2}{3}} |\pi^0 n
\ra + \sqrt{\frac{1}{3}} | \pi^- p \ra,
 \ee
and
 \be
\label{eq:1901102}
  |K \Sigma; \frac{3}{2}, -\frac{1}{2} \ra =
\sqrt{\frac{2}{3}} |K^0 \Sigma^0 \ra + \sqrt{\frac{1}{3}} |K^+
\Sigma^- \ra, \ee respectively.
According to Eqs.~(\ref{eq:1901101}) and (\ref{eq:1901102}), the
coefficients $C_{ij}$ in the isospin space can be calculated and
listed in Table~\ref{table:coef_S0I32}. Since the coefficients are
all negative, the Weinberg-Tomozawa contact interaction between the
pseudoscalar meson and the baryon octet is repulsive in the sector
of isospin $I=3/2$ and strangeness $S=0$.
Even if the correction from the $s-$ channel and $u-$ channel
interaction is taken into account, as is shown in
Fig.~\ref{fig:tchi32s0}, the total potential of the pseudoscalar
meson and the baryon octet is still repulsive.
Thus no resonance state could be generated in the S-wave
approximation.

\begin{table}[htbp]
\begin{tabular}{c|cc}
\hline\hline
 $C_{ij}$         & $\pi N$ &   $K \Sigma$    \\
\hline
$\pi N$            &$-1$ & $-1$  \\
$K \Sigma$         &$$ & $-1$    \\
\hline \hline
\end{tabular}
\caption{The coefficients $C_{ij}$ in the pseudoscalar meson and
baryon octet interaction with isospin $I=\frac{3}{2}$ and
strangeness $S=0$, $C_{ji}=C_{ij}$.} \label{table:coef_S0I32}
\end{table}

\section{Summary}
\label{sect:summary}

In this work, the interaction of the pseudoscalar meson and the
baryon octet is studied within a nonlinear realized Lagrangian. The
$s-$, $u-$ channel potentials and the Weinberg-Tomozawa contact
interaction are obtained when the three-momenta of the particles in
the initial and final states are neglected in the S-wave
approximation.

In the sector of isospin $I=1/2$ and strangeness $S=0$, a resonance
state is generated dynamically by solving the Bethe-Salpeter
equation, which might be regarded as counterparts of the $N(1535)$
 particle listed in the PDG data.
We find the hidden strange channels, such as $\eta N$, $K \Lambda$
and $K \Sigma$, play an important role in the generation of the
resonance state when the Bethe-Salpeter equation is solved in the
unitary coupled-channel approximation.
The coupling constants of this resonance state to different channels
are calculated, and it is found that it couples strongly to the
hidden strange channels.

\begin{acknowledgments}
Bao-Xi Sun would like to thank Han-Qing Zheng and Yu-Fei Wang for
useful discussions.
\end{acknowledgments}

\newpage

\section*{Appendix: Subtraction constants, pole position and couplings of the resonance state to different channels}
\begin{table}[htbp]
\begin{tabular}{c|ccccccccccccc}
\hline\hline
   $n$          & $a_{\pi N}$ & $a_{\eta N}$ &   $a_{K \Lambda}$ &   $a_{K \Sigma}$ & $Pole~position(MeV)$ & $g_{\pi N}$ & $|g_{\pi N}|$ & $g_{\eta N}$ & $|g_{\eta N}|$ &  $g_{K \Lambda}$ & $|g_{K \Lambda}|$ &   $g_{K \Sigma}$
   &   $|g_{K \Sigma}|$  \\
\hline
  1  &  -3.2  &  -1.7  &  -3.2  &  -3.2  &
 1518-46i  &
  -3+  1i  &    4   &
 -65+ 25i  &   70  &
  41+  0i  &   41  &
  94-27i  &   98 \\
  2  &  -3.2  &  -1.4  &  -3.2  &  -3.2  &
 1530-58i  &
  -4+  1i  &    4  &
 -66+ 28i  &   72  &
  40+  3i  &   40  &
  95-27i  &   99 \\
  3  &  -2.9  &  -2.3  &  -3.2  &  -2.9  &
 1520-41i  &
  -3+  5i  &    7  &
 -62+ 21i  &   66  &
  47+  0i  &   47  &
  96-24i  &   99 \\
  4  &  -2.9  &  -2.0  &  -3.2  &  -2.9  &
 1532-51i  &
  -4+  5i  &    7  &
 -64+ 22i  &   68  &
  45+  3i  &   45  &
  97-24i  &  100 \\
  5  &  -2.9  &  -1.7  &  -3.2  &  -3.2  &
 1518-46i  &
  -3+  2i  &    4  &
 -65+ 25i  &   70  &
  42+  0i  &   42  &
  94-27i  &   98 \\
  6  &  -2.9  &  -1.4  &  -3.2  &  -3.2  &
 1530-58i  &
  -4+  1i  &    4  &
 -66+ 28i  &   72  &
  40+  4i  &   40  &
  95-27i  &   99 \\
  7  &  -2.6  &  -2.3  &  -3.2  &  -2.9  &
 1520-41i  &
  -2+  6i  &    7  &
 -62+ 21i  &   66  &
  47+  1i  &   47  &
  96-24i  &   99 \\
  8  &  -2.6  &  -2.3  &  -2.9  &  -2.9  &
 1527-41i  &
  -2+  4i  &    5  &
 -60+ 19i  &   64  &
  48+  1i  &   48  &
  95-21i  &   98 \\
  9  &  -2.6  &  -2.0  &  -3.2  &  -2.9  &
 1532-51i  &
  -3+  6i  &    7  &
 -64+ 22i  &   68  &
  45+  4i  &   45  &
  97-24i  &  100 \\
 10  &  -2.6  &  -1.7  &  -3.2  &  -3.2  &
 1518-46i  &
  -3+  2i  &    4  &
 -65+ 25i  &   70  &
  42+  0i  &   42  &
  94-27i  &   98 \\
 11  &  -2.6  &  -1.4  &  -3.2  &  -3.2  &
 1530-59i  &
  -4+  2i  &    5  &
 -67+ 28i  &   73  &
  40+  4i  &   40  &
  95-27i  &   99 \\
 12  &  -2.3  &  -2.3  &  -3.2  &  -2.9  &
 1520-41i  &
  -2+  7i  &    8  &
 -62+ 21i  &   66  &
  47+  1i  &   47  &
  96-25i  &   99  \\
 13  &  -2.3  &  -2.0  &  -3.2  &  -2.9  &
 1532-51i  &
  -2+  6i  &    8  &
 -64+ 23i  &   68  &
  45+  4i  &   46  &
  97-24i  &  100 \\
 14  &  -2.3  &  -1.7  &  -3.2  &  -3.2  &
 1518-46i  &
  -2+  3i  &    4  &
 -65+ 25i  &   70  &
  42+  1i  &   42  &
  94-27i  &   98 \\
 15  &  -2.3  &  -1.4  &  -3.2  &  -3.2  &
 1530-59i  &
  -3+  2i  &    5  &
 -67+ 28i  &   73  &
  40+  4i  &   40  &
  95-27i  &   99 \\
 16  &  -2.0  &  -2.6  &  -3.2  &  -2.6  &
 1535-47i  &
   0+ 11i  &   12  &
 -62+ 18i  &   65  &
  51+  5i  &   52  &
  97-22i  &  100 \\
 17  &  -2.0  &  -2.3  &  -3.2  &  -2.9  &
 1520-41i  &
   0+  7i  &    8  &
 -62+ 21i  &   66  &
  48+  1i  &   48  &
  95-25i  &   99 \\
 18  &  -2.0  &  -2.0  &  -3.2  &  -2.9  &
 1531-51i  &
  -1+  7i  &    8  &
 -64+ 23i  &   68  &
  46+  4i  &   46  &
  96-24i  &  100 \\
 19  &  -2.0  &  -1.7  &  -3.2  &  -3.2  &
 1518-46i  &
  -2+  3i  &    5  &
 -65+ 25i  &   70  &
  42+  1i  &   42  &
  94-27i  &   98 \\
 20  &  -2.0  &  -1.4  &  -3.2  &  -3.2  &
 1530-59i  &
  -3+  3i  &    5  &
 -67+ 28i  &   73  &
  40+  4i  &   40  &
  95-27i  &   99 \\
 21  &  -1.7  &  -2.6  &  -3.2  &  -2.6  &
 1534-47i  &
  -1-12i  &   12  &
 -62+ 18i  &   65  &
  52+  5i  &   52  &
  97-22i  &  100 \\
 22  &  -1.7  &  -2.3  &  -3.2  &  -2.9  &
 1519-41i  &
  -1 -8i  &    8  &
 -62+ 21i  &   66  &
  48+  1i  &   48  &
  95-25i  &   99 \\
 23  &  -1.7  &  -2.0  &  -3.2  &  -2.9  &
 1531-51i  &
   0+  7i  &    8  &
 -64+ 23i  &   69  &
  46+  4i  &   46  &
  96-24i  &   99 \\
 24  &  -1.7  &  -1.7  &  -3.2  &  -3.2  &
 1518-46i  &
  -1+  3i  &    5  &
 -65+ 25i  &   70  &
  42+  1i  &   42  &
  94-27i  &   98 \\
 25  &  -1.7  &  -1.4  &  -3.2  &  -3.2  &
 1530-59i  &
  -2+  3i  &    5  &
 -67+ 28i  &   73  &
  40+  5i  &   41  &
  95-27i  &   99 \\
 26  &  -1.4  &  -2.6  &  -3.2  &  -2.6  &
 1534-47i  &
  -3-11i  &   11  &
 -62+ 19i  &   65  &
  53+  4i  &   53  &
  97-22i  &   99 \\
 27  &  -1.4  &  -2.3  &  -3.2  &  -2.9  &
 1519-41i  &
  -2 -8i  &    8  &
 -62+ 21i  &   66  &
  49+  1i  &   49  &
  95-25i  &   98 \\
 28  &  -1.4  &  -2.0  &  -3.2  &  -2.9  &
 1531-51i  &
  -1 -8i  &    8  &
 -64+ 23i  &   69  &
  47+  4i  &   47  &
  96-25i  &   99 \\
 29  &  -1.4  &  -1.4  &  -3.2  &  -3.2  &
 1530-59i  &
  -2+  4i  &    5  &
 -67+ 28i  &   73  &
  41+  5i  &   41  &
  95-28i  &   99 \\
 30  &  -1.1  &  -2.6  &  -3.2  &  -2.6  &
 1534-46i  &
  -4-10i  &   11  &
 -62+ 19i  &   65  &
  53+  4i  &   53  &
  96-22i  &   99 \\
 31  &  -1.1  &  -2.3  &  -3.2  &  -2.9  &
 1519-41i  &
  -3 -7i  &    8  &
 -62+ 21i  &   66  &
  49+  0i  &   49  &
  95-25i  &   98 \\
 32  &  -1.1  &  -2.0  &  -3.2  &  -2.9  &
 1531-51i  &
  -2 -7i  &    7  &
 -64+ 23i  &   69  &
  47+  4i  &   47  &
  96-25i  &   99 \\
 33  &  -1.1  &  -1.4  &  -3.2  &  -3.2  &
 1530-59i  &
  -1+  4i  &    5  &
 -67+ 28i  &   73  &
  41+  5i  &   41  &
  95-28i  &   99 \\
 34  &  -0.8  &  -2.6  &  -3.2  &  -2.6  &
 1533-46i  &
  -5 -9i  &   10  &
 -62+ 19i  &   65  &
  53+  3i  &   53  &
  96-22i  &   98 \\
 35  &  -0.8  &  -2.0  &  -3.2  &  -2.9  &
 1531-51i  &
  -2 -7i  &    7  &
 -64+ 23i  &   69  &
  47+  3i  &   48  &
  95-24i  &   99 \\
 36  &  -0.8  &  -1.4  &  -3.2  &  -3.2  &
 1530-59i  &
   0+  3i  &    4  &
 -67+ 28i  &   73  &
  41+  5i  &   41  &
  95-28i  &   99 \\
 37  &  -0.5  &  -2.6  &  -3.2  &  -2.6  &
 1533-46i  &
  -6 -8i  &   10  &
 -62+ 19i  &   65  &
  54+  3i  &   54  &
  95-21i  &   98 \\
 38  &  -0.5  &  -2.0  &  -3.2  &  -2.9  &
 1531-51i  &
  -3 -6i  &    7  &
 -64+ 23i  &   69  &
  48+  3i  &   48  &
  95-24i  &   98 \\
 39  &  -0.5  &  -1.4  &  -3.2  &  -3.2  &
 1530-59i  &
   0 -4i  &    4  &
 -67+ 28i  &   73  &
  41+  4i  &   42  &
  94-28i  &   99 \\
\hline \hline
\end{tabular}
\caption{The values of the subtraction constant $a_{\pi N}$,
$a_{\eta N}$, $a_{K \Lambda}$, $a_{K \Sigma}$, the pole position on the complex energy plane of $\sqrt{s}$ and  couplings of the resonance state to different channels. The regularization
scale $\mu=630$MeV is fixed in the loop function in Eq.~(\ref{eq:Our G result}). }
\label{table:subtract}
\end{table}

\end{document}